\documentclass[portrait]{article}

\usepackage{times}				
\usepackage{url}                

\usepackage{graphicx}			

\usepackage{amssymb}

\usepackage[authoryear]{natbib}	
								
\usepackage[portrait,margin=0.5in]{geometry}

\usepackage{hyperref}
\usepackage{gensymb}
\usepackage{amsmath}

\title{Exobodies in Our Back Yard:
\\Science from Missions to Nearby Interstellar Objects
\\Science White Paper submitted to the 2023-2032 Planetary Science and Astrobiology Decadal Survey}
\author{T. Marshall Eubanks
\\(tme@space-initiatives.com)
\\Space Initiatives Inc
\\Jean Schneider 
\\Paris Observatory-LUTh
\\ Andreas M. Hein
\\ Adam Hibberd
\\ Robert Kennedy
\\ Institute for Interstellar Studies US
}

\begin{document}
\maketitle

\shortcites{MACRO-2002-a}
\pagestyle{plain}

\begin{abstract}
The recent discovery of the first confirmed Interstellar Objects (ISOs)
passing through the Solar System 
on clearly hyperbolic objects
opens the potential for near term ISO missions, either to the two known objects, or to similar objects found in the future.
Such ISOs are the only exobodies we have a chance of accessing directly in the near future.
This White Paper focuses on the science possible from \textit{in situ} spacecraft exploration 
of nearby ISOs. 
Such spacecraft missions are technically possible now and are suitable potential missions in the period covered by the 2023-2032 Decadal Survey.
Spacecraft missions can determine the structure and the chemical and isotopic composition of ISO in a close flyby coupled with a small sub-probe impactor and either a mass spectrometer or a high resolution UV spectrometer; this technology will also be useful for fast missions to TransNeptune Objects (TNOs) and long period comets. 
ISO exploration holds the potential of providing considerable improvements in our knowledge of galactic evolution, of planetary formation, and of the cycling of astrobiologically important materials through the galaxy. 
\end{abstract}

\section{Introduction}

The existence of Interstellar Objects (ISOs) visiting the Solar System has been predicted for many years (e.g. \cite{Sekanina-1976-a,Stern-1990-a}). We live in an interesting epoch where two of them have been found (1I/'Oumuamua) and (2I/Borisov) and signs of other visitors have been proposed \citep{Siraj-Loeb-2018-a,Froncisz-et-al-2020-a}. In addition, there are prospects to detect one or several of them per year with the Vera Rubin Observatory (VRT) starting in 2022 \citep{Trilling-et-al-2017-a}. Ground-based telescopes will not be able to give reliable answers to questions about the origin, chemical composition, mineral structure, size, shape of these interstellar visitors - that will require 
\textit{in situ} exploration.

During its history the solar system must have  ejected vast numbers of planetesimals, comets and asteroids into interstellar space.
Recent discoveries have shown that exoplanetary systems are common;  if they have similar histories, there should thus be a sizable  population of small protoplanetary ISO  \citep{Engelhardt-et-al-2017-a}. In addition, stars can eject dust and gas and planetary debris \textit{after} their time on the main sequence, and it is even possible that small bodies form directly in interstellar space \citep{Eubanks-2019-a}.  Although 
spacecraft, radar and visual detection arrays definitely observe streams of dust-sized interstellar micrometeorites  \citep{Strub-et-al-2015-a,Baggaley-2000-a}, and although there appears to be a population of nomadic interstellar planets \citep{Sumi-et-al-2011-a,Strigari-et-al-2012-a,Clanton-Gaudi-2017-a}, until recently there was no firm evidence for even a single galactic asteroid or comet coming into the solar system from interstellar space \citep{Engelhardt-et-al-2017-a}.

In October 2017, the first known interstellar object (ISO), named 1I/'Oumuamua, was detected by
the \textit{Pan-STARRS1} survey, and then observed and tracked  as it was moving through the solar system at a heliocentric velocity of
$\sim$50 km/s.  1I/'Oumuamua was detected after $\sim$3.5 years of \textit{Pan-STARRS1}  observing in its current survey mode.  \cite{Do-et-al-2018-a} used this to calculate an upper limit to the space density, n, of similar sized ISOs of
\begin{equation}
\mathrm{n}\ \lesssim\ 0.2\ \mathrm{AU}^{-3}\ .
\label{eq:number-density}
\end{equation}
indicating these objects may pass through the inner solar system more or less continuously and that targeted searches \citep{Eubanks-2019-b} or more sensitive survey instruments such as the VRT \citep{Trilling-et-al-2017-a} may be able to find numerous targets for future missions. 

\begin{figure}[!ht]
\begin{center}
\includegraphics[scale=0.4]{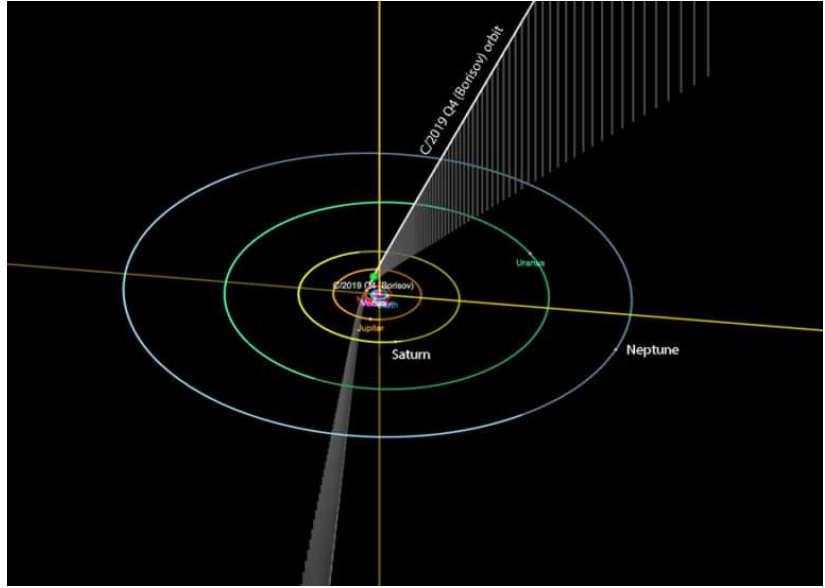}
\end{center}
\caption{The trajectory of 
2I/Borisov, discovered by the amateur astronomer Gennadiy  Borisov on August 30 2019 (M.P.E.C. 2019-R106). Its minimum heliocentric distance was 1.96 $\pm$ 0.04 AU  with an orbital eccentricity e = 3.155. As of July 15 2020 its distance to the Earth is 4.74 AU and in mid-2030 it will be $\sim$74 AU. Recent observations suggest that it is now fragmented 
\citep{Jewitt-et-al-2020-a}.
}
\label{fig:Borisov_trajectory}
\end{figure}

The discovery of the interstellar comet 2I/Borisov on August 30, 2019, \citep{Guzik-et-al-2019-a} provides a second ISO available for study in its passage through our solar system (see Figure \ref{fig:Borisov_trajectory}). 2I/Borisov is an active object superficially similar to a small solar system comet \citep{Jewitt-and-Luu-2019-a},  has  an estimated nucleus
diameter of order 1 km \citep{Jewitt-et-al-2019-a}, and was rapidly recognized to be on a hyperbolic orbit.
2I/Borisov was discovered  at an R magnitude of 17.8 in the Northern sky, and reached a peak apparent magnitude of about 15 near perihelion. Based on data from the IAU Minor Planet Center database, long period comets have been discovered at similar magnitudes since at least Comet Kohoutek (C/1973 E1) in 1973, 48 years before 2I/Borisov was discovered. This strongly suggests that interstellar comets as large and active as 2I/Borisov are actually rare enough that it may be decades before another such target is found.  

The transit of ISOs through the solar system offers the opportunity to discover and study 
ISOs at close range directly by \textit{in situ} spacecraft exploration \citep{Hein-et-al-2017-a,Hibberd-et-al-2019-b}.   ISOs likely form very far from the solar  system in both space and time, possibly in a completely different part of the Milky Way Galaxy. Exploration of ISOs holds the promise of determining the 
creation mechanisms of small planetesimals during the formation of solar-type stellar systems, during the formation of other types of stellar systems (such as M-dwarf systems), and also of observing objects formed in the demise of main sequence stellar systems and ejected during the creation of white dwarf and neutron star systems.

\section{Dynamical Streams as Proxies for ISO Birth Systems}

Stellar perturbations make it hard to predict the detailed galactic trajectories of 
ISOs over intervals much longer than a few million years \citep{Zhang-2017-a}, 
$<$ 10$^{-3}$ of the history of the galaxy, and no firm association with a nearby star has been claimed for either 1I/'Oumuamua or 2I/Borisov \citep{Bailer-Jones-et-al-2019-a,Hallatt-Wiegert-et-al-2019-a}.
The difference between the ability to retrodict ISO orbits and their likely nomadic lifetimes of many gigayears means that  it is unlikely that we will ever find the natal star system of a passing ISO. However,  
association of ISOs with stellar dynamic streams can serve as a proxy to provide bounds on the age, metallicity,  chemical composition and even origin of the source stellar system \citep{Eubanks-2019-a,Eubanks-2019-b}. 

\begin{figure}[!ht]
\begin{center}
\includegraphics[scale=1.0]{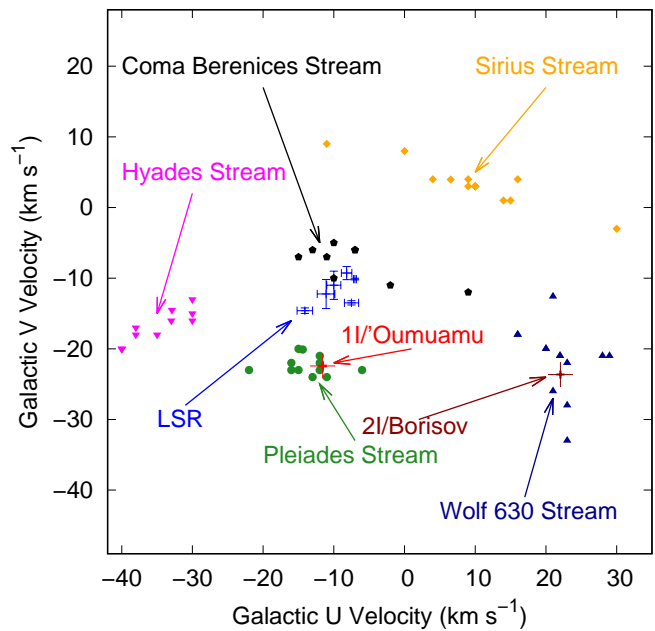}
\end{center}
\caption{The galactocentric U and V components of velocity for two ISOs, the Local Standard of Rest (LSR) and five local dynamical streams, from
\cite{Eubanks-2019-c}. The 1I and 2I incoming velocities are near the centroids of the determinations of the velocities of the Pleiades and Wolf 630 streams, respectively, and these objects presumably were members of these streams before their encounter with the solar system. The stream velocity estimates are from  \citep{Kushniruk-et-al-2017-a}, supplemented by \citep{Chereul-et-al-1998-b,Liang-et-al-2017-a,Gaia-et-al-2018-b}.
The 1I and LSR velocites are as shown in \citep{Eubanks-2019-b}, and the 2I velocity estimate is from the ephemeris of Bill Grey / Project Pluto.
}
\label{fig:Velocity-Scatter}
\end{figure}

The Milky Way's velocity fields are highly non-uniform \citep{Ramos-et-al-2018-a},
with a substantial fraction of the stars in the solar neighborhood being concentrated 
in dynamical streams (also known as stellar associations or moving groups) moving in the galaxy
(see, e.g., \citep{Kushniruk-et-al-2017-a,Gaia-et-al-2018-b}). 
Figure \ref{fig:Velocity-Scatter} shows  that 
the incoming 1I/'Oumuamua velocity vector ``at infinity'' (\textbf{v}$_{\infty}$) is close to the local velocity of the  Pleiades dynamical stream; this asteroid was presumably a member of this stream
\citep{Feng-Jones-2018-a,Eubanks-2019-a,Eubanks-2019-b}. Likewise, the \textbf{v}$_{\infty}$ of 2I/Borisov
is close to the motion of the Wolf 630 dynamical stream in all three velocity components, suggesting that it was (before its interaction with the solar system) a member of that stream \citep{Eubanks-2019-c}.

Associating an ISO with a dynamical stream allows the stars
in the stream to be used as proxies for natal stellar system of the ISO.  
Data from a spectroscopic survey of stars with distances $<$ 250 pc
\citep{Quillen-et-al-2018-b} reveals that the Wolf 630
possesses a slightly higher than solar metallicity ([Fe/H] $>$ 0.2 dex).
Antoja \textit{et al} 
\citep{Antoja-et-al-2008-a} determined the kinematics of 
over 16,000 nearby stars with known ages; the Wolf 630 
stream (their moving group 17) has in their data almost no stars younger than 
 0.5 Gyr, and are mostly 2 - 8 Gyr old. The lower bound of this range agrees well
 with an earlier estimate for the stellar age of the Wolf 630 stream of 2.7 $\pm$ 0.5 Gyr
 \citep{Bubar-King-2010-a}. These data suggest that 2I/Borisov may be roughly the age of the solar system.  By contrast, the Pleiades stream   consists stars with ages between roughly 10$^{7}$ and 10$^{9}$ years old  \citep{Chereul-et-al-1998-b,Famaey-et-al-2008-a,Bovy-Hogg-2009-a}, suggesting that 
1I/'Oumuamua is a relatively young object. 
 
 \section{ISOs and the Chemical Evolution of the Galaxy}

The chemical evolution of the galaxy is dominated by elements created by 
slow neutron capture (the s-process)  in  Asymptotic Giant Branch (AGB) stars, 
\citep{Busso-et-al-1999-a},
and by the rapid processes present
in explosive nucleosynthesis in Supernova (SN) \citep{Yoshida-2007-a}.
ISOs will contain these materials in different concentration than for solar system material;
determining the isotopic and elemental composition of ISOs will provide direct information about the formation mechanisms of these objects. 
The composition of presolar grains in primitive meteorites 
shows that they predominately come from these two sources \citep{Davis-2011-a}; half 
of all the elements heavier than iron are thought to result from the s-process \citep{Herwig-2005-a}.
AGB material is thus very
important in the chemical evolution of the galaxy, and its presence in stardust seems to 
permeate the galactic disk \citep{Gail-et-al-2009-a}. \cite{Eubanks-2019-c} described how ISOs created in the demise of stellar systems (i.e., Post Main Sequence Objects) could thus be very common in the galaxy, and  could form a substantial part of the ISO flux passing through the solar system.

\section{Missions to Interstellar Objects}

The feasibility of reaching interstellar objects at close encounter with the Sun has been assessed in
\cite{Seligman-Laughlin-2018-a}, with specific missions to the interstellar objects 1I/'Oumuamua and 2I/Borisov being presented by 
\cite{Hein-et-al-2017-a}  and \cite{Hibberd-et-al-2019-b}. Such missions are feasible 
with more massive spacecraft (100 kg or larger) using existing rockets and technologies
either if they can be initiated around the time of perihelion passage, e.g. by a Comet Interceptor type mission \citep{Schwamb-et-al-2020-a}, or by using a combination of planetary flybys and solar Oberth maneuvers (rocket accelerations at high speed close to the Sun) to overtake 
the ISO as it retreats from the Sun \citep{Hein-et-al-2017-a,Hein-et-al-2019-a}.
Small sailcraft could also possibly explore ISOs \citep{Turyshev-et-al-2020-a}, and could also act as very valuable adjuncts to larger missions, as they
could serve as widely distributed precursor probes to find small and dark ISOs years or decades after their last observation from Earth.  \cite{Hibberd-Hein-2020-a} demonstrate that a mission to 1I/'Oumuamua would be feasible, using a GW-scale beaming infrastructure and a series of 1-100 kg probes. At 300 km/s, 1I/'Oumaumua could be reached within 1-2 years, even if the probe were launched as late as 2030. Due to the faintness of some ISOs (e.g. 1I/'Oumuamua), the capability of launching multiple spacecraft increases the likelihood that at least one of the spacecraft might obtain useful data, even if other spacecraft fly past the ISO at a too large distance. Nevertheless, the mass of the spacecraft will be 1-100 kg or even below for lower beaming power. As a consequence, only a limited suite of instruments can be carried on an individual spacecraft, compared to a probe such as New Horizons. Launching individual spacecraft with different instruments, aggregating incoming data, and using these probes as a common spacecraft platform would effectively lead to a more efficient exploration of ISOs than via a single, monolithic spacecraft.

\section{Science With Extreme Hypervelocity Impacts} 
\label{App:Impact-Science}

No matter what technique is used to reach an passing ISO, the relative spacecraft-ISO velocities at the time of flyby will be large, probably $>>$ 10 km s$^{-1}$, and there thus is little prospect of the spacecraft entering orbit or performing a soft landing on the ISO. The ISO mission science will have to be executed in a fast flyby, and the
science return would be enhanced with a sub-probe impactor.

Hypervelocity impacts have  relative velocities  $\gtrsim$ 3 - 5  km s$^{-1}$,  speeds where the strength of materials is negligible compared to the impact forces. At these velocities,  the impact energy / atom controls the immediate (prompt) response to the impact. As an example, it  takes about 0.4 eV / molecule to vaporize water, while water molecules have a binding energy of $\sim$4.4 eV, and the first ionization of both Hydrogen and Oxygen requires $\sim$13.6 eV.  There is thus a profound difference between an impact at 5 km s$^{-1}$ ($\sim$2 eV / water molecule), which will produce mostly superheated steam, and an impact at 15 km s$^{-1}$ ($\sim$18 eV / water molecule), which will produce an ionized plasma. 

Impacts with energies / atom $\gtrsim$ 20 eV can thus be usefully described as extreme hypervelocity impacts, with dramatic changes in the impact physics,  corresponding to temperatures $\gtrsim$ 2 $\times$ 10$^{5}$ K and higher.
The resulting ionized prompt plumes will  produce radiation at Extreme UltraViolet (EUV, 10 to 120 nm) and soft X-ray (XUV, 0.1 to 10 nm) wavelengths, depending on the collision energy, which can be used to investigate the physics of the impact and the composition of the impacted bodies.

\subsection{The Physics of Hypervelocity Impacts}

An ISO mission with a subprobe impactor would conduct an extreme hypervelocity impact creating a highly ionized prompt plume with energies up to $\sim$ 300 eV, or $\sim$3 million K. 
A small impactor will not change the velocity of the impacted asteroid by more than a few mm s$^{-1}$, and so the ISO fixed frame can be viewed as an inertial frame, and the atomic constituents of the ISO can be viewed as initially at rest in that frame. Assume, as a first order approximation, a non-relativistic head-on elastic collision between an atom in the impactor, of mass m$_{i}$ and initial velocity 
v$_{i_{in}}$, and an atom in the asteroid, with mass m$_{A}$ and zero velocity in the asteroid rest frame. Then the post-collision velocities are given by
\begin{equation}
v_{i_{out}} = \frac{m_{i} - m_{A}}{m_{i} + m_{A}}\ v_{i_{in}}
\label{eq:elastic-1-D-i}
\end{equation}
and
\begin{equation}
v_{A_{out}} = \frac{2\ m_{i}}{m_{i} + m_{A}}\ v_{i_{in}}  ;
\label{eq:elastic-1-D-A}
\end{equation}
Atoms with small atomic mass compared to those in the impactor will receive a large velocity change (up to twice the impact velocity) but a relatively small fraction of the  incoming atom's Kinetic Energy (KE), while more massive atoms will have a smaller velocity change, but can absorb more of the incoming atom's KE. The energies considered in this paper are not large enough to initiate nuclear reactions, but it is reasonable to assume that cohesive and molecular bonds will be broken, and electrons removed, up to the maximum amount of energy available.

\subsection{Comparison of Hypervelocity Impact Missions}

Table \ref{table:impact-missions} provides information about  proposed and flown hypervelocity impact missions. 
At an impact velocity of $\sim$18 km s$^{-1}$ a hypothetical 1I flyby mission with a sub-probe impacting on 1I/'Oumuamua (henceforth called a 1I mission) would produce more highly ionized ejecta 
than did either the Deep Impact (DI) mission, which struck the comet Tempel 1 with an impact velocity of 10.3 km s$^{-1}$ \citep{AHearn-et-al-2005-a} or the proposed DART impact of the moon of the asteroid Didymos \citep{Cheng-et-al-2015-b}, with a planned impact velocity of $\sim$6.5 km s$^{-1}$. The nominal 1I Mission impactor is a 7 kg rod, 1 m long and 1 cm in radius,  of isotopically pure $^{190}$Pt (this isotope was chosen as it is commercially available).

Table \ref{table:impact-energies} provides the collision energy imparted to some of the elements commonly found in asteroids and comets. This Table shows that, while the DART impact will not have enough energy to ionize most elements, DI should have singly ionized much of the material in its prompt plume. The 1I Mission would deeply ionize the fast plume material (although only Hydrogen and Helium should be fully stripped of their electrons and thus only those elements should produce K, or Lyman alpha, spectral lines). 

The DI impactor included a 178.4 kg copper mass, which, given the density of copper (8960 kg m$^{-3}$), would be a sphere with a radius of   168.1  mm. 
A prompt ``hot plume'' and a longer duration ``ejecta plume'' were observed by  the main DI spacecraft, with the ejecta plume containing much more matter (a mass ratio of $\sim$3 $\times$ 10$^{4}$) but only $\sim$20\% of the prompt plume energy \citep{Groussin-et-al-2010-a}. (The mass and energy budget totals have uncertainties of order 50\%.) The prompt DI ejecta plume included a cloud of  incandescent liquid droplets which dominated the optical emission from the plume for $\sim$ 420 msec. The droplet cloud, with a center of mass motion of $\sim$7.9 km s$^{-1}$ and an internal expansion rate of 1.7 km s$^{-1}$, are thought to have been a 4000 kg cloud of  $\sim$160 $\mu$m diameter particles of cooling silica melt. Scaling this to the 1I mission impacts indicates that it would be hazardous to attempt a direct sampling of the plume by a spacecraft. 


\begin{table} 
\caption{Impact Velocities and Energies for Planned and Proposed Hypervelocity impact missions. The column marked $\lambda$ is the Compton wavelength of the impact energy.   For Deep Impact and DART, the composition and the Energy / Atom is for the predominant element in the Impactor. The detailed elemental breakdown of the DART impactor is not known, but the spacecraft frame will be an Aluminum alloy, and that is assumed to be the predominant element. The DI impactor mass includes 6.5 kg of  unused hydrazine fuel; it is assumed that the ``49\% copper'' listed in \citep{AHearn-et-al-2005-a} refers to the dry mass. }
\begin{tabular}{c c c c c r r}
 \noalign{\smallskip}
\cline{1-7}
Mission          &  \multicolumn{2}{c}{Impactor} & V$_{impact}$  & Energy & Energy & $\lambda$   \\
                        &  Mass & Composition & & & / Atom & \\
DART   \citep{Cheng-et-al-2016-b}          &  $\sim$300 kg & mostly Al & $\sim$6.5 km s$^{-1}$  & 6.3 GJ &   5.9 eV & 209.9 nm  \\
DI          \citep{AHearn-et-al-2005-b} & 370.5 kg &  49\% Cu  & 10.3   km s$^{-1}$ & 19.7 GJ &  34.9 eV & 35.5 nm\\
1I-Mission                & 10 kg & 7 kg $^{190}$Pt & $\sim$18.2 km s$^{-1}$  & 1.65 GJ & 335.7 eV & 3.7 nm \\
\cline{1-7}
\end{tabular}
\label{table:impact-missions}
\end{table}

Table \ref{table:impact-energies} shows the ionization of common elements in the DART, DI and 1I Mission impact plumes. 
While the 1I Mission impact should provide enough kinetic energy to directly ionize deuterium but not hydrogen (in this context, $^{1}$H or protium) secondary collisions and a fast equipartition of energy should provide enough kinetic energy to fully ionize Hydrogen and Helium, and thus generate Hydrogen Lyman emission between 121.6 and 91.18 nm. Deuterium Lyman Alpha lines are shifted bluewards by $\sim$0.03 nm from the protium lines, corresponding to a velocity shift of $\sim$80 km s$^{-1}$; these lines should thus be separable given adequate spectral resolution.

The EUV spectrum of the 1I Mission impact plume  should span the range of $\sim$5 to 122 nm. The ALICE EUV instrument on the \textit{New Horizons} spacecraft has a spectral pass band of 42 - 187 nm \citep{Stern-et-al-2008-a}. The Alice instrument would be able to observe Hydrogen and Deuterium L$_{\alpha}$, but a redesigned UV telescope with fast optics would be required to observe the rapidly varying EUV signatures expected from a 1I mission-type impact.

\begin{table} 
\caption{Energy changes for head-on elastic impact by the main constituent of the impactor for various missions; for elements more massive than Helium, the terrestrial isotopic distribution is assumed. The ionization refers to the energy produced in a primary impact;  equipartition of energy will make more energy available to low mass elements from secondary impacts and should result in the full ionization of Hydrogen and Helium in the 1I Mission plume.}
\begin{tabular}{c r c r c r c}
 \noalign{\smallskip}
\cline{1-7}
\\
Element          & \multicolumn{2}{c}{DART}        & \multicolumn{2}{c}{DI}  & \multicolumn{2}{c}{1I Mission} \\
                       & $\Delta$KE & Ionization & $\Delta$KE & Ionization & $\Delta$KE & Ionization  \\
$^{1}$H   & 0.8 eV & - & 2.1 eV & - & 6.8 eV & - \\
$^{2}$H   & 1.5 eV & - & 4.2 eV & - & 13.5 eV & - \\
$^{3}$He & 2.1 eV & - & 6.1 eV & - & 20.0 eV & - \\
$^{4}$He & 2.7 eV & - & 7.8 eV & - & 26.3 eV & I \\
Li               & 3.8 eV & - & 12.4 eV & I & 44.4 eV & I \\
C               & 5.0 eV & - & 18.8 eV & I & 73.1 eV & IV \\
O               & 5.5 eV & - & 22.6 eV & I & 93.7 eV & IV\\
Mg             & 5.9 eV & - & 28.0 eV & II & 131.8 eV & IV \\
Si               & 5.9 eV & - & 30.1 eV & II & 147.2 eV & IV \\
Fe              & 5.1 eV & - & 35.5 eV & III & 231.8 eV & VIII \\
\cline{1-7}
\end{tabular}
\label{table:impact-energies}
\end{table}

\section{ISO Missions and Interstellar Medium}

At about 100 AU, the Solar Wind becomes subsonic and beyond 120 AU it encounters the Local InterStellar Medium (LISM). While the Voyager 1 and 2 probes are currently exploring this region, several fast probes in different directions and at different distances would provide a 3D map of the heliosphere up to around 200 AU. In particular the region opposite to the direction of the solar wind with respect to the LISM \citep{Katushkina-et-al-2014-a}, i.e. at RA DEC = 20H, 28 deg (the Cygni constellation), would reveal a possible plasma magnetotail of the Solar wind in the LISM. Whether this magnetotail exists is still a matter of debate, which can likely only be resolved by direct \textit{in situ} observation, which could be done as part of a ISO chase mission.

The IBEX Ribbon is a compelling possible target for ISO missions. 
The IBEX Ribbon is an unexpected feature discovered by IBEX in the flux of energetic neutral hydrogen. It  covers a wide swath of the sky roughly 20$\degree$ by 75$\degree$ and is hypothesized to be due to interactions between the heliosphere and the LISM \citep{Fichtner-et-al-2014-a}. The parallax of the ribbon has been estimated to be 140 $\substack{+84 \\ -38}$ AU \citep{Zirnstein-2019-a}
rendering the entire Ribbon region accessible to direct \textit{in situ} exploration by  
ISO chase missions. Exploration of an extended region over 100 AU in length would certainly benefit from a series of probes sampling many locations; this can be done relatively inexpensively by a fleet of fast sailcraft. 
We recommend that ground based astronomical resources be utilized to find TNOs in the foreground of the Ribbon \citep{Trilling-et-al-2018-c}; flyby exploration of such objects would be a useful complement to a Ribbon exploration campaign.


\bibliographystyle{plainnat}


\end{document}